# Enhanced continuous aerosol optical depth (AOD) estimation using geostationary satellite data: focusing on nighttime AOD over East Asia

Sanghyeon Song[1, 2], Yoojin Kang[2], Jungho Im[2,3*]

[1]Department of Civil Engineering, McGill University, Montréal, QC H3A 0C3, Canada

[2]Department of Civil, Urban, Earth, and Environment Engineering, Ulsan National Institute of Science and Technology, Ulsan, 44919, Republic of Korea

[3]Department of Environmental Resources Engineering, State University of New York, College of Environmental Science and Forestry, Syracuse, NY 13210, USA

\* Corresponding author

Jungho Im (ersgis@unist.ac.kr)

Office) +82-52-217-2824, Fax) +82-52-217-2849

# Abstract

Continuous aerosol monitoring in East Asia is essential due to the massive aerosol emissions from natural and anthropogenic sources. Geostationary satellites enable continuous aerosol monitoring; however, the observation is limited to the daytime. This study proposed machine learning-based models to estimate daytime and nighttime aerosol optical depth (AOD) in East Asia using a geostationary satellite, Geo-KOMPSAT-2A (GK-2A). The input variables for the machine learning models include the brightness temperature (BT) and top-of-atmosphere (TOA) reflectance from GK-2A, meteorological and geographical data, and auxiliary variables. The two models that used different combinations of GK-2A variables were proposed and compared: the all-day BT model, which estimates AOD during both day and night using BT variables, and the daytime TOA model, which estimates AOD during the day using TOA reflectance variables as well. The estimated AODs by the models were validated with ground-based AOD data from the Aerosol Robotic Network (AERONET) by 10-fold cross-validation and hold-out validation methods. The performance of the daytime TOA model was slightly higher than the all-day BT model during the day ($R^2$ = 0.80–0.82, root mean square error (RMSE) = 0.107–0.116 for the all-day BT model, $R^2$ = 0.83, RMSE = 0.098 for the daytime TOA model). The SHapley Additive exPlanations (SHAP) analysis showed that total precipitable water content and seasonality contributed the most for both proposed models. BT differences and TOA reflectance variables were identified as the next most contributing variables for the all-day BT and daytime TOA models. The spatiotemporal distributions of estimated AODs from the proposed models show similar patterns compared with other AOD products. A time series comparison at a test station demonstrated that the estimated AOD of the proposed models was consistent with the AERONET AOD.

# 1. Introduction

Atmospheric aerosols are liquid droplets and solid particles suspending in air. Aerosols affect the Earth's radiative energy budget and further climate through their interaction with radiation and clouds (Boucher, Randall, Artaxo, ..., & 2013, 2013). In addition, aerosols have an effect on human health and the ecosystem, including biodiversity in vegetation and agricultural productivity (Kang, Kim, Kang, Cho, & Im, 2022; Knippertz et al., 2015; Shiraiwa et al., 2017). Therefore, it is essential to monitor the spatial distribution and temporal variation of atmospheric aerosols and to understand their optical and physical properties. Especially in East Asia, due to extensive natural emissions (e.g., Asian dust) and massive artificial emissions from industrialization and urbanization, continuous aerosol monitoring is necessary (W. Li et al., 2016; Maki et al., 2019).

Aerosol optical depth (AOD) is a parameter that represents the attenuation of radiation by aerosols in a total atmospheric column as a function of wavelength, and it is used to measure the quantity of aerosols in the atmosphere. AOD is monitored via ground- and satellite-based observations. A global ground-based observation network, the Aerosol Robotic Network (AERONET), provides long-term, continuous ground-based observations of aerosol properties with high quality so that they are used as reference data for calibrating and validating satellite-based AODs (M. Choi et al., 2019; Giles et al., 2019; Kang et al., 2021). However, the ground-based AOD observations are point-based measurements that do not provide observation over large areas, with unbalanced distribution of stations, and require costly station maintenance (Wei, Chang, Bai, & Gao, 2019; Yang, Xu, & Yu, 2020).

Satellite observations, on the other hand, can provide vast spatial coverage for regional or global aerosol monitoring through Earth-observing sensors such as the Moderate Resolution Imaging Spectroradiometer (MODIS) and the Advanced Himawari Instrument (AHI). The majority of satellite-based AOD retrieval algorithms rely on prior knowledge of the relationship between aerosol properties and surface reflectance based on radiative transfer models, which often simplify these variables or utilize pre-calculated look-up tables (LUTs) for computational efficiency. However, it can introduce uncertainty caused by errors from calibration, aerosol model selection, cloud masking, incorrect assumptions on surface reflectance building LUTs, and interpolation errors by non-linearity (Grey, North,

& Los, 2006; Kittaka, Winker, Vaughan, Omar, & Remer, 2011; Levy et al., 2010; Remer et al., 2005; She, Zhang, Li, de Leeuw, & Huang, 2020).

Recently, machine learning has been utilized to supplement the existing satellite-based AOD retrieval algorithms to improve accuracy (A. Chen et al., 2023; Y. Chen et al., 2022; Kang et al., 2022; Lops, Pouyaei, Choi, …, & 2021, 2021; She et al., 2020; Z. Shi et al., 2022; Su, Laszlo, Li, Wei, & Kalluri, 2020; Yeom et al., 2022). Data-driven machine learning models can simulate complicated non-linear relationships between satellite observations and AOD from extensive previous data. Machine learning models can directly estimate AOD from spatiotemporally heterogeneous input factors without being constrained by LUTs, and consequently, they can reduce uncertainties that originated from LUTs. Machine learning methods are often heavily dependent on the selection of training data and parameters, and are usually effective only within the scope of variables used during training; therefore, it is crucial to be aware of this regard when using those methods (Huttunen et al., 2016).

The majority of satellite-based AOD retrieval algorithms have been developed focusing on daytime observations. It is challenging to retrieve AOD at night due to the absence of a reliable standard light source (Kocifaj & Bará, 2022). Consequently, a limited number of methods for retrieving AOD at night have been studied until now. The satellite-borne active sensor, Cloud-Aerosol Lidar and Infrared Pathfinder Satellite Observations (CALIPSO), provides a global aerosol profile, including at night; however, it has limited spatiotemporal coverage due to the narrow 5-km swath of the instrument and 16-day revisit interval (X. Ma, Bartlett, Harmon, & Yu, 2013; D. M. Winker et al., 2010; David M. Winker et al., 2009). Several studies have attempted to retrieve AOD during the nighttime using moonlight or artificial light from existing satellite sensors, such as the Visible Infrared Imaging Radiometer Suite (VIIRS) Day/Night band (J. Wang et al., 2020; J. Zhang et al., 2019; Zhou et al., 2021). Some earlier studies employed artificial neural networks to retrieve nighttime dust AOD using MODIS infrared (IR) brightness temperature (BT) (S. S. Lee et al., 2017; Sang-Sam & Sohn, 2012). While these algorithms can retrieve coarse-mode dust AOD from a polar orbit satellite, the idea of using IR channels to retrieve nighttime AOD based on machine learning is noteworthy. However, these observation methods are limited to a narrow swath of area, and real-time AOD monitoring is not available due to the long revisit time of polar orbit satellites. To understand the diurnal variation of

AOD and facilitate real-time monitoring of aerosol transport, geostationary satellite-retrieved nighttime AOD should be utilized in conjunction with daytime observations. Notwithstanding the critical nature of nighttime aerosol monitoring, an operational algorithm that encompasses nighttime over a vast area (e.g., East Asia) for continuous AOD monitoring does not yet exist.

In this study, we employed machine learning methods to estimate all-time AODs from geostationary satellite Geostationary-Korean Multi-Purpose Satellite-2A (GK-2A) data using ground-based AERONET AOD as reference data. In particular, we focused on estimating nighttime AOD using data-driven methods and ensuring that the performance is similar and consistent to the daytime AOD estimation performance. To the best of our knowledge, this is the first study focusing on estimating the nighttime AOD using geostationary satellite data and thereby retrieving the continuous diurnal variation of AOD over East Asia from space.

## 2. Study domain and data

### 2.1 Study domain

The study area is East Asia, which includes China, Japan, Russia, Mongolia, and the northern part of the Indochina Peninsula. It is clipped from the observation area of East Asia provided by the GK-2A satellite, considering the distribution of the AERONET stations (Fig. 1). The period of this study is from August 2019 to August 2021.

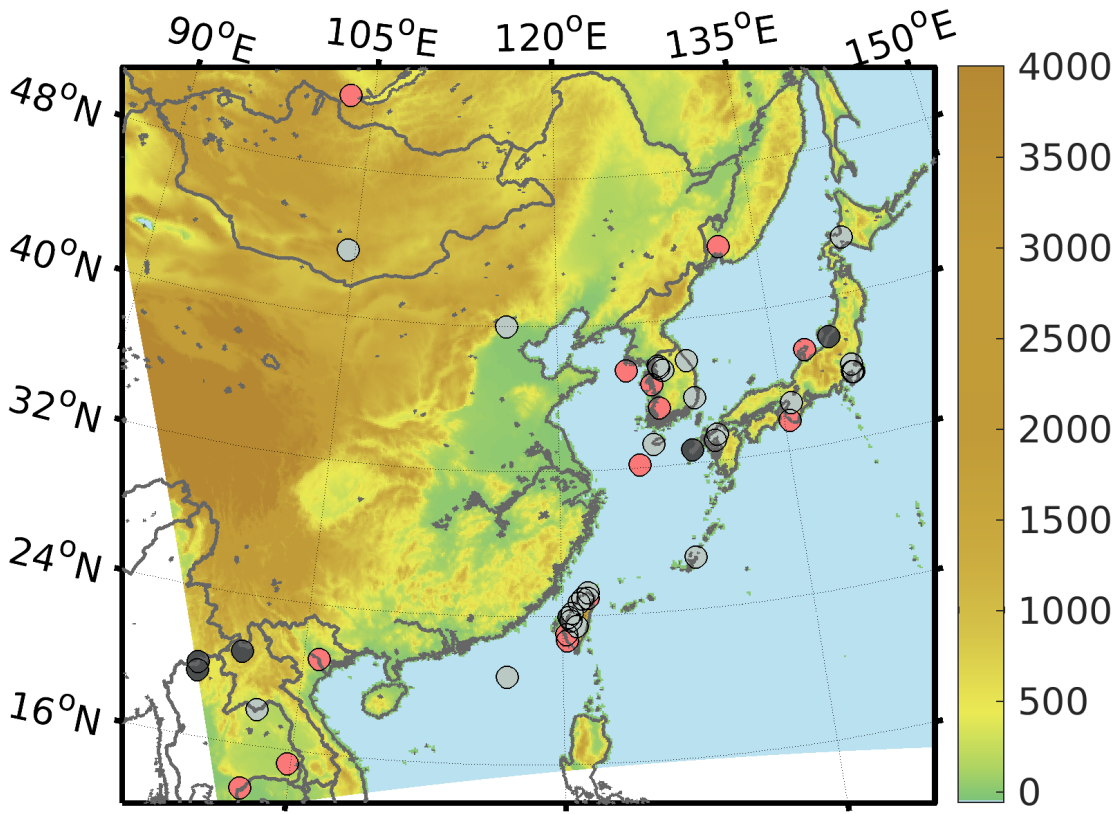

**Fig. 1.** The study area with ground-based aerosol monitoring AERONET stations. The stations that provide only Solar AOD and Lunar AOD are shown as red and black, each, and the stations that provide both Solar and Lunar AOD are shown as grey. The background is surface elevation from JAXA/ALOS World 3D-30m (AW3D30) Digital Surface Model (DSM).

### 2.2 Data

#### 2.2.1 Ground-based AOD measurements

AERONET Version 3 AOD products were used as ground-based reference data. AOD products from AERONET, a global network of ground-based aerosol observations, have been used as a benchmark to evaluate the performance of satellite-based aerosol monitoring (Falah et al., 2021; Giles et al., 2019; Y. R. Shi et al., 2021). AERONET provides long-term aerosol data with high temporal resolution at three levels of data quality: Level 1.0 (unscreened), Level 1.5 (cloud-cleared), and Level 2.0 (quality-assured). While AERONET provides aerosol information using the Direct Sun Algorithm with the three levels of data quality at daytime, AERONET measures nighttime AOD using the Moon as a light source by the Direct Moon Algorithm (provisional) with two quality levels of Level 1.0 and Level 1.5 since

2015 (Barreto et al., 2016; Zhou et al., 2021). In this study, AERONET Version 3 Level 2.0 solar AOD products from 39 stations were used as reference data for daytime, and the Level 1.5 lunar AOD products from 29 stations were used for nighttime. Twenty-two of these stations provide both solar and lunar AOD measurements (Fig. 1).

### 2.2.2 GK-2A/AMI

GK-2A is a geostationary meteorological satellite of Korea, launched by the Korea Meteorological Administration (KMA) in December 2018. The Advanced Meteorological Imager (AMI) onboard GK-2A has 16 channels, consisting of 6 reflective channels—4 visible and 2 near-infrared (NIR) channels—and 10 emissive channels—4 mid-wavelength and 6 long-wavelength infrared (IR) channels (S. Lee & Choi, 2021) (Table S1). While the reflective channels collect data only during the day when solar radiation exists, the emissive channels produce observations during both day and night. GK-2A Level 1B (L1B) data were preprocessed into TOA reflectance and BT and used as input variables for machine learning models. The details of preprocessing are provided in Section 3.1.

GK-2A Level 2 (L2) meteorological products were also used as input variables. As atmospheric water vapor is one of the essential parameters for aerosol retrieval, total precipitable water (TPW), the vertical distribution of atmospheric water vapor, was used as an input variable to consider the influence of atmospheric water on the aerosol optical property (Gui et al., 2017; Vijayakumar & Devara, 2012). GK-2A provides TPW products every 10 minutes with a spatial resolution of 6 km. In addition, in satellite-based aerosol monitoring, the retrieval of aerosol properties in cloudy areas is hindered. Thus, cloudy regions are masked out using the GK-2A cloud detection product, which is produced every 10 minutes over East Asia with a spatial resolution of 2 km. The cloud detection product of GK-2A is provided with three classes: 0 for cloud, 1 for probable cloud, and 2 for clear sky. The pixels with a value of 0 for the cloud mask were eliminated from the AOD estimation. In addition, to consider the effect of partial clouds on aerosols, the cloud detection product, except for the value of 0, was used as an input variable. Similar to the cloud, pixels with fog were also masked out using the GK-2A fog detection product to prevent the model from overestimating the AOD in the area with fog, especially sea fog. As with the cloud detection product, GK-2A fog detection products are provided every 10 minutes with a spatial resolution of 2 km in East Asia. All input variables used in this study are summarized in Table 1.

### 2.2.3 Meteorological data

Various meteorological data generated from a numerical weather prediction model were used to estimate AOD considering the relationships between the meteorological data and aerosols (B. Chen et al., 2021; Kang et al., 2022; S. Park et al., 2020; She et al., 2020; Tariq, Nawaz, Ul-Haq, & Mehmood, 2021). It is known that some meteorological variables are related to the transportation of aerosols (e.g., wind speed and components, surface pressure, PBLH), the formation, growth, and decomposition of aerosols (e.g., air temperature, RH), and optical extinction by aerosols (i.e., visibility) (Guo et al., 2023; Kang et al., 2022; J. Li et al., 2020; Z. Li et al., 2017; Tariq et al., 2021; H. Wang et al., 2019; S. Zhang, Wu, Fan, Yang, & Zhao, 2020; X. Zhang, Chu, Wang, & Zhang, 2018).

The Regional Data Assimilation and Prediction System (RDAPS) is one of the numerical weather forecasting models operated by KMA based on the United Model (UM). RDAPS provides meteorological forecast data four times per day (i.e., 00, 06, 12, and 18 UTC) with a spatial resolution of 12 km, adopting the initial and boundary conditions from the Global Data Assimilation and Prediction System (GDAPS) (Kang et al., 2022; S. Park et al., 2020). In this study, wind speed, u- and v-components of wind, surface pressure, temperature, dew point temperature, relative humidity, visibility, and planetary boundary layer height (PBLH) were used as input variables.

### 2.2.4 Geographical and auxiliary data

The land cover and Normalized Difference Vegetation Index (NDVI) products from MODIS were used to consider geographical factors. The MODIS land cover (MCD12Q1) product was used to consider the effect of land cover on aerosol distribution. Among the five land cover classification schemes of the MCD12Q1, we used the International Geosphere-Biosphere Programme (IGBP) land cover scheme. It provides 17 classes of land cover classification annually derived from MODIS Terra and Aqua reflectance data with a spatial resolution of 500 m. In this study, the ratio of a specific land cover type in a moving-window of 12 km was calculated using the 500 m land cover data. The NDVI product from MODIS Aqua (MYD13A2), which provides NDVI at a spatial resolution of 1 km every 16 days, was also used to consider the effect of vegetation on aerosols.

Solar zenith angle (SZA), day of year (DOY), and hour of day (HOD) were used as auxiliary variables. SZA was calculated from the longitude and latitude of the grid and time using the NOAA Solar Calculator (*https://gml.noaa.gov/grad/solcalc/*). DOY and HOD were used to reflect the seasonal and diurnal dependences of AOD in the model. SZA is related to the calculation of scattering angles, which affect AOD retrieval (M. Zhang et al., 2020).

### 2.2.5 Existing AOD data for comparison

The AODs at 550 nm from AHI Yonsei Aerosol Retrieval (YAER) and Copernicus Atmosphere Monitoring Service (CAMS) were used as the comparative products. The AHI YAER algorithm is an aerosol retrieval algorithm using observation from AHI onboard the Himawari-8 satellite and calculates aerosol optical properties at 550 nm with 6 km spatial resolution in East Asia (5° S–50° N, 110° E–150° E). Aerosol products are calculated from AHI L1B TOA reflectance and BT through the inversion process using a LUT, pre-calculated based on a radiative transfer model (Lim, Choi, Kim, Kasai, & Chan, 2018).

CAMS global atmospheric composition forecasts provide AOD products not only for daytime but also for nighttime over a large area. CAMS is the global air quality data assimilation model incorporating satellite data and produces global forecasts for more than 50 chemical species and 7 types of aerosol composition (ECMWF, 2021). It provides forecasts for the next 120 hours at 1-hour intervals twice a day, during both daytime and nighttime, with a horizontal resolution of approximately 40 km.

**Table 1.** Input variables used for the machine learning-based models in this study. The name of all variables and their abbreviation is in the bracket. The spatial resolution is the original spatial resolution of each variable. All variables are resampled to a spatial resolution of 12 km. The detailed preprocessing steps of variables are discussed in Section 3.1.

| Data Source | Variables | Spatial Resolution |
|---|---|---|
| GK-2A L1B | TOA reflectance differences with recent 30-day minimum: ch01_diff, ch02_diff, ch03_diff, ch04_diff, ch05_diff, ch06_diff | 0.5 – 2 km (Table S1) |
| | BT difference between channels: ch07-ch13 (BTD1) ch13-ch15 (BTD2) ch11-ch13 (BTD3) BT ratios between channels: ch13/ch11 (BTR1) ch13/ch15 (BTR2) BT differences with recent 30-day maximum: | 2 km |

| | ch11_diff, ch13_diff, ch14_diff, ch15_diff | |
|---|---|---|
| GK-2A L2 | Cloud | 2 km |
| | Fog | 2 km |
| | Total Precipitable Water (TPW) | 6 km |
| UM-RDAPS | Temperature (Temp) | 12 km |
| | Dew point Temperature (Dew) | |
| | Relative Humidity (RH) | |
| | Wind Speed (WS) | |
| | U-component of wind (U-wind) | |
| | V-component of wind (V-wind) | |
| | Visibility | |
| | Planetary Boundary Layer Height (PBLH) | |
| | Surface Pressure (Psrf) | |
| MODIS | Land Cover ratio of Vegetation (LC_veg) | 0.5 km |
| | Land Cover ratio of Water (LC_wat) | |
| | Land Cover ratio of Urban (LC_urb) | |
| | Normalized Difference Vegetation Index (NDVI) | 1 km |
| Auxiliary data | Solar Zenith Angle (SZA) | - |
| | Day of Year (DOY) | - |
| | Hour or Day (HOD) | - |

## 3. Methods

### 3.1 Data preprocessing

All input data were resampled to a 12 km spatial resolution to match the RDAPS longitude and latitude grids, which have the largest grid size among the input variables. For the data with different spatial resolutions, GK-2A L1B and L2 data, land cover ratios, and NDVI were upscaled to the spatial resolution of RDAPS data by averaging pixels by block and re-gridded to the RDAPS grid using bilinear spatial interpolation.

For comparison with other AOD products, AERONET AOD products at wavelengths of 500 nm and 675 nm were interpolated to AOD at 550 nm using the following equation (Bibi et al., 2015; Y. Chen et al., 2022; Kang et al., 2022)

$$AOD_{550} = AOD_{500} \left(\frac{550}{500}\right)^{-\alpha} \quad (1)$$

$$\alpha = -\frac{\ln\frac{AOD_{500}}{AOD_{675}}}{\ln\frac{500}{675}} \quad (2)$$

where α is the Ångström exponent of 500-675 nm. The converted AERONET AOD observations at 550 nm were averaged for 60 min, which is 30 min before and after every hour. The latitude and longitude of AERONET stations were matched with the coordinates of the nearest RDAPS grid.

Many of the current satellite-based AOD retrieval algorithms utilize TOA reflectance to retrieve AOD through the physical relationship between TOA reflectance and AOD. As the reflectance observed by a satellite sensor consists of contributions from the surface and the atmosphere, several methods have been developed to separate the contributions of each to the observed reflectance (Kolmonen, Sogacheva, Virtanen, de Leeuw, & Kulmala, 2016; Yang et al., 2020). Recent studies have utilized the minimum reflectance technique, which uses the minimum TOA reflectance value as the surface reflectance, to better separate the contributions (M. Choi et al., 2016, 2019; Kang et al., 2022). In this study, the differences between the observed TOA reflectance and its minimum within the recent 30 days of 6 reflective channels (i.e., ch01 to ch06) were used as input variables to account for the contribution of atmospheric aerosols to the TOA reflectance. BT data were preprocessed into 3 BT differences (BTDs), 2 BT radios (BTRs), and 4 BT differences with their recent 30-day maximum in 4 IR channels (i.e., ch11, ch13, ch14, and ch15). BT differences and ratios between two wavelength channels serve as indicators of aerosol type and are widely used in aerosol detection based on radiative transfer models (Jee, Lee, Lee, & Zo, 2020; S. S. Park et al., 2014). As atmospheric aerosols reduce BT, the maximum BT in the recent 30 days was considered the background (Jee et al., 2020; Kim et al., 2014).

RDAPS meteorological data, which provided analysis fields with intervals of 6 hours, were temporally interpolated to 1-hour intervals through linear interpolation. The MODIS land cover classification data were converted into ratios for each land cover type. The land cover ratios of cropland, grasslands, and forest types are summed up as a 'vegetation' type. For days for which NDVI data were not available, the data from the closest day were used. DOY was converted to have values between -1 (for the winter season) and 1 (for the summer season) over 365 days of the year using the sine function (Stolwijk, Straatman, and Zielhuis, 1999; Park et al., 2022). In a similar way, HOD was also transformed to have values between -1 and 1 over 24 hours of a day using the sine function (S. Park et al., 2022).

Then, cloud masking and fog masking were applied to the GK-2A L1B channel data. If more than 50% of the pixels in a 12 km window were masked out due to cloud contamination, the corresponding upscaled pixel was also masked out. For fog masking, pixels classified as 'fog' and 'probably fog' classes of the GK-2A fog detection product were masked out.

The oversampling approach was applied to the training data to avoid biased modeling due to the sparse distribution of AERONET and the data imbalance problem. The oversampling approach adopted in this study is based on the assumption that neighboring pixels have similar AOD values (S. Park et al., 2019). Oversampling was performed only for samples with an AOD of 0.3 or more in the three stations that consistently experienced high AOD levels (Fig. S1). The adjacent 8 pixels centered on a pixel were randomly perturbed within ±5% of the AOD of the central pixel. The direction of perturbation was determined by the TPW considering the positive correlation between TPW and AOD (i.e., if the TPW of an adjacent pixel is smaller than the TPW of the center pixel, a lower perturbed value was allocated to the nearby pixel).

### 3.2 Estimation of AOD

In this study, we developed two machine learning models for estimating AOD with different combinations of input variables. The first model, named the "all-day BT model," uses BT-derived variables from GK-2A L1B data along with auxiliary data. Another model with additional TOA reflectance as input variables, named the "daytime TOA model," was also proposed and compared with the all-day BT model. In this study, daytime and nighttime were defined as when SZA was below 85 degrees and above 85 degrees, respectively. The auxiliary input variables, except for GK-2A L1B data, were the same for the two models. The process flow for estimating daytime and nighttime AOD using machine learning-based models is illustrated in Fig. 2.

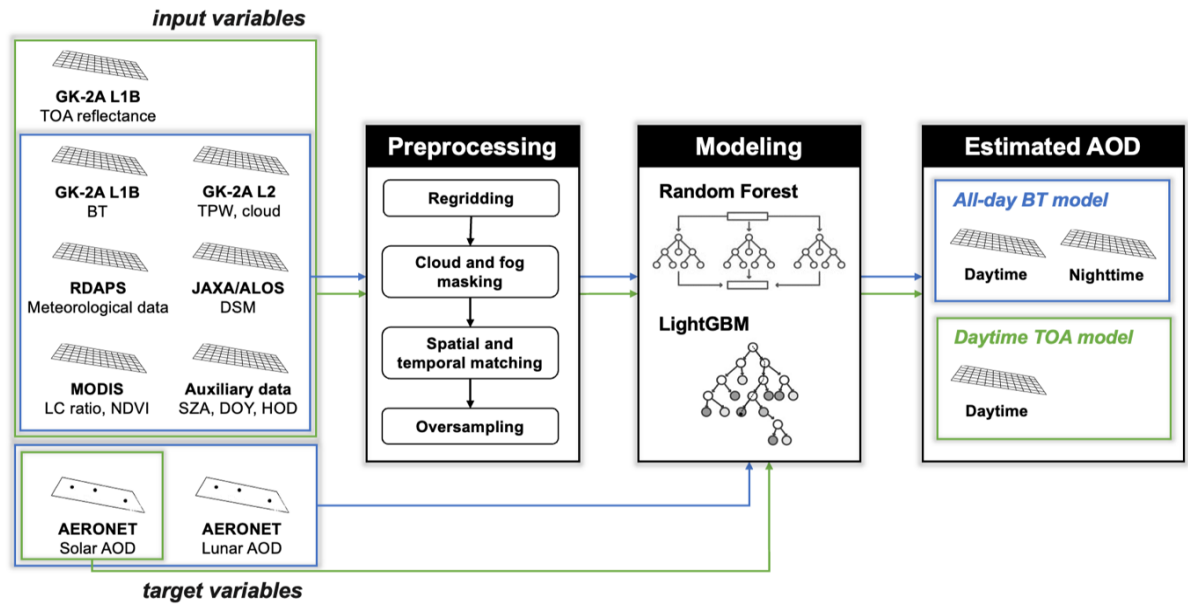

**Fig. 2.** Process flow for the estimation of AOD proposed in this study. The all-day BT model uses the same input variables during both daytime and nighttime, and the daytime TOA model additionally uses GK-2A L1B TOA reflectance as input variables during daytime.

### 3.3 Machine learning models

In this study, two machine learning algorithms were used for estimating daytime and nighttime AOD: random forest (RF) and light gradient boosting machine (LGBM). RF is a tree-based bagging ensemble algorithm that aggregates numerous randomly generated weak learners to build a strong learner. RF generates a multitude of decision trees using a bootstrap approach for samples and variables and aggregates the outputs from each tree to obtain the final prediction (Breiman, 2001). RF has been widely used in recent studies on air quality monitoring and prediction as it can quantify the non-linear relationships between atmospheric and geographical parameters in remote sensing (Mao et al., 2021). The RF model was built and optimized using the scikit-learn package in Python 3.8.13 environment (Pedregosa FABIANPEDREGOSA et al., 2011). The hyperparameters for RF were optimized using the Python library "hyperopt," based on the Bayesian optimization algorithm (Bergstra, Yamins, & Cox, 2013). The optimized hyperparameters were set as in Table S2.

LGBM is an ensemble learning algorithm based on gradient-boosting decision trees. LGBM uses gradient-based one-side sampling and exclusive feature bundling to efficiently deal with a large amount of data (Ke et al., 2017; J. Ma, Zhang, Xu, & Yu, 2022). LGBM builds decision trees by splitting training data in the direction of decreasing errors and obtains the prediction by optimally combining decision trees (Kang et al., 2022). Compared to other gradient-boosting algorithms that grow trees horizontally (i.e., level-wise tree growth), LGBM grows trees vertically (i.e., leaf-wise tree growth) by choosing the leaf with the maximum delta loss to grow. The leaf-wise tree growth is computationally faster than the level-wise tree growth. The LGBM model for this study was implemented using the "LGBM" package in the Python 3.8.8 environment. The hyperparameters for LGBM were optimized using the Python library "hyperopt," the same as for RF. The hyperparameter optimization processes were conducted every time the model was run to prevent overfitting or underfitting issues for the leaf-wise growth algorithm (Microsoft Corporation, 2023; J. Zhang et al., 2019). The optimized hyperparameters of LGBM are summarized in Table S2.

### 3.4 Evaluation and analysis methods

The proposed models were evaluated using two types of hold-out validations (spatially and temporally) and a 10-fold cross-validation (CV). For the hold-out validations, a specific ground-based station and days were left out to evaluate the trained models (i.e., hold-out test sets). Data from one AEORNET station (KORUS_UNIST_ULSAN, South Korea, Fig. S2) among the 46 AERONET stations was selected as the test dataset for spatial hold-out validation, and data with DOY in multiples of five from the remaining data were left out as the test dataset for temporal hold-out validation. In addition, a random 10-fold CV (RDCV) was performed using the remaining data to demonstrate the robustness of the machine learning models given the limited number of samples (H. Choi, Park, Kang, Im, & Song, 2023; Kang et al., 2022; She et al., 2020).

The model performance of AOD estimation was quantitatively assessed using four accuracy metrics: the coefficient of determination [$R^2$, (3)], root mean square error [RMSE, (4)], relative RMSE [rRMSE, (5)], index of agreement [IOA, (6)], and the fraction within, above and below the expected error in percentage [EE, (7)].

$$R^2 = \frac{\sum_{i=1}^{n}(\hat{y}_i - \bar{y})^2}{\sum_{i=1}^{n}(y_i - \bar{y})^2} \tag{3}$$

$$RMSE = \sqrt{\frac{\sum_{i=1}^{n}(\hat{y}_i - y_i)^2}{n}} \tag{4}$$

$$rRMSE = \frac{\sqrt{\frac{\sum_{i=1}^{n}(\hat{y}_i - y_i)^2}{n}}}{\bar{y}} \times 100\ [\%] \tag{5}$$

$$IOA = 1 - \frac{\sqrt{\sum_{i=1}^{n}(y_i - x_i)^2}}{\sum_{i=1}^{n}(|x_i - \bar{y}| + |y_i - \bar{y}|)^2} \tag{6}$$

$$EE = \pm(0.15 \times AOD_{AERONET} + 0.05) \times 100\ [\%] \tag{7}$$

where y and ŷ are observed and predicted data, and $\bar{y}$ is the mean of the observed data. IOA is a standardized measure of model accuracy with a range from 0 to 1; a higher IOA indicates higher model accuracy (Willmott, 2013). EE is often used to assess the uncertainty of satellite-based AOD retrieval with ground-based AOD, and the higher the percentage of samples within EE, the higher the accuracy of the model. In this study, the EE of the MODIS Dark Target AOD algorithm with AERONET AOD over land was applied for evaluation (Levy et al., 2013).

The contribution and interactions of input variables to the models were analyzed using the Shapley Additive Explanations (SHAP) method (Lundberg, Allen, & Lee, 2017). SHAP quantifies the contribution of each feature in a prediction in terms of both direction and magnitude by calculating the mean of all the predicted values and the contributions of each feature to the predictions based on game theory (Kang et al., 2021; S. Park et al., 2022). In this study, model interpretation using SHAP is implemented with the "shap" package in Python 3.8.13 environment (Lundberg et al., 2020).

4. Results and discussion

    4.1 Model performance

Overall, the LGBM model performed better than the RF model based on the all-day RDCV results (Table 2; Fig. 3a). In addition, the slope of the line of best fit for the LGBM model is closer to 1 than that of the RF model for both daytime and nighttime. The LGBM model underestimated AOD less than the RF model, which is supported by the fact that larger fractions of samples within EE for the LGBM model than the RF model. The difference of the performances between RF and LGBM was based on their ability of addressing different aspects of error. LGBM is a boosting algorithm that aims to reduce the bias, whereas RF is a bagging algorithm that focuses on decreasing the variance (Fan, Xiao, Sun, Zhang, & Xu, 2022). As shown in Fig. 3, the LGBM model resulted in the estimated AOD matching more closely with the reference AOD throughout all ranges of AOD due to its characteristics of reducing bias. The RF model, on the other hand, caused higher errors in high AOD values due to the small sample size for high AOD and its characteristic of reducing variance.

The temporal hold-out test demonstrated superior performance to the spatial test across all time periods for both machine learning models. It is more difficult to estimate AOD of an unseen location than to estimate the AOD of a day not used as model training. The model performance exhibited a decline when data from a specific station was excluded from training and used for testing, owing to discrepancies in climate, surface reflectance, and aerosol optical properties across AERONET stations and their uneven distribution (She et al., 2020).

The addition of TOA reflectance improved daytime performance for both the RF and LGBM models (Table 3). In comparison to the all-day BT model, the underestimation trend was mitigated in the daytime TOA model (Fig. 3). In addition, the daytime TOA model performed better than the all-day BT model, as evidenced by $R^2$ increases of 0.63 to 0.72 for the temporal test and 0.51 to 0.66 for the spatial test, respectively. Similar to the RDCV results, the spatial and temporal tests revealed that the daytime model underestimated AODs less than the all-day BT model (Fig. S3).

**Table 2.** Performance assessment results of the all-day BT model for two machine learning models by each model validation method for all-day (all), daytime, and nighttime. The model validation methods include random cross-validation (RDCV) and two types of hold-out validation (temporal and spatial tests). The estimated AODs from the all-day BT model were separated into daytime and nighttime results by SZA to compare the model performance between the daytime and nighttime.

|     |               |       | R² | RMSE | rRMSE | IOA |
|-----|---------------|-------|------|-------|---------|------|
| RF  | RDCV          | all   | 0.75 | 0.129 | 51.88 % | 0.90 |
|     |               | day   | 0.75 | 0.127 | 51.48 % | 0.89 |
|     |               | night | 0.78 | 0.136 | 53.45 % | 0.91 |
|     | Temporal test | all   | 0.58 | 0.147 | 57.14 % | 0.83 |
|     |               | day   | 0.58 | 0.142 | 55.42 % | 0.83 |
|     |               | night | 0.57 | 0.167 | 63.79 % | 0.83 |
|     | Spatial test  | all   | 0.45 | 0.118 | 54.11 % | 0.73 |
|     |               | day   | 0.47 | 0.120 | 54.62 % | 0.74 |
|     |               | night | 0.36 | 0.109 | 51.99 % | 0.71 |
| LGBM | RDCV         | all   | **0.81** | **0.109** | **44.03 %** | **0.94** |
|     |               | day   | **0.80** | **0.107** | **43.59 %** | **0.94** |
|     |               | night | **0.82** | **0.116** | **45.71 %** | **0.94** |
|     | Temporal test | all   | 0.61 | 0.137 | 55.69 % | 0.87 |
|     |               | day   | 0.63 | 0.132 | 54.57 % | 0.87 |
|     |               | night | 0.63 | 0.155 | 59.63 % | 0.87 |
|     | Spatial test  | all   | 0.50 | 0.112 | 54.57 % | 0.79 |
|     |               | day   | 0.51 | 0.114 | 54.36 % | 0.79 |
|     |               | night | 0.45 | 0.104 | 55.26 % | 0.79 |

**Table 3.** Performance assessment results of the daytime TOA model with TOA reflectance differences with recent 30-day minimum as additional input variables for two machine learning models.

|      |               | R² | RMSE | rRMSE | IOA |
|------|---------------|------|-------|---------|------|
| RF   | RDCV          | 0.76 | 0.125 | 50.38 % | 0.90 |
|      | Temporal test | 0.64 | 0.132 | 51.07 % | 0.87 |
|      | Spatial test  | 0.58 | 0.108 | 51.19 % | 0.81 |
| LGBM | RDCV          | **0.83** | **0.100** | **40.79 %** | **0.95** |
|      | Temporal test | 0.72 | 0.114 | 47.75 % | 0.91 |
|      | Spatial test  | 0.66 | 0.098 | 49.29 % | 0.86 |

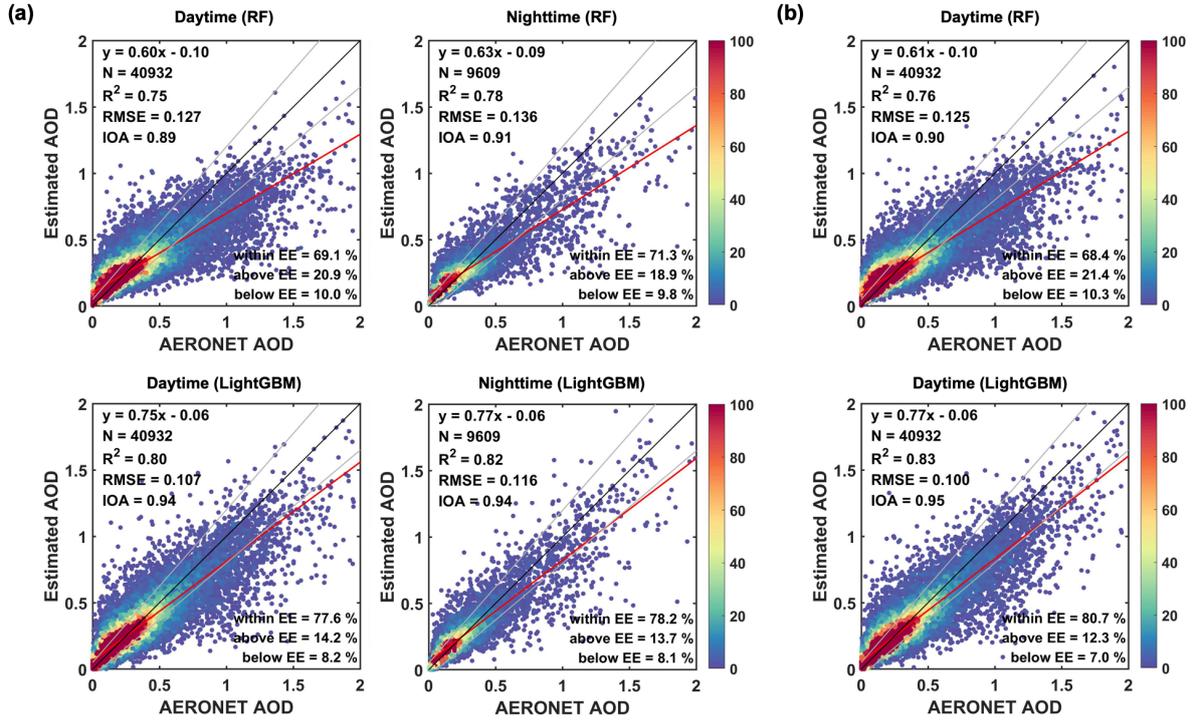

**Fig. 3.** Scatter plots of random 10-fold cross-validation (RDCV) for (a) the all-day BT and (b) the daytime TOA models based on RF and LGBM. The x-axis is the reference AERONET AOD, and the y-axis is the estimated AOD. The black line is an identity line, the red line is a line of best fit, and the grey thin lines are expected errors (EE). The color of the points indicates the number of samples.

### 4.2 Model interpretation

Fig. 4 shows the summary plots of SHAP values for both the daytime and all-day models based on the best-performing LGBM. For both models, TPW and DOY contributed significantly to the AOD estimation. There is a significant relationship between TPW and AOD because a higher precipitable water content in the atmosphere contributes to the growth of aerosols (Anoruo, 2023; Ogunjobi & Awoleye, 2019). In addition, as AOD has a strong seasonal pattern in the study area, DOY was effectively used to represent the seasonality of AOD (Kang et al., 2022; S. Park et al., 2020).

In both models, NDVI and LC ratios also greatly contributed to the estimation of AOD. NDVI has been often used in AOD retrievals to reduce a systematic bias at low AOD caused by bright surfaces (M. Choi et al., 2016; Kang et al., 2022; Levy et al., 2013). Furthermore, higher urban ratio values were generally related to higher AOD, indicating that the urban ratio could represent anthropogenic AOD in urban areas to some extent (Yu, Wong, & Liu, 2023). Overall, meteorological variables were less important than other groups of variables. Among meteorological variables, for the all-day BT model, dew point temperature and surface pressure moderately contributed to the AOD estimation: the higher the dew point temperature, the higher the AOD, and the lower the surface pressure, the lower the AOD.

In the all-day BT model, the BT difference variables, especially BTD1 (ch07–ch13) and BTD2 (ch13–ch15), highly contributed to estimating AOD (Fig. 4a). These two BT difference variables have usually been used for dust aerosol or cloud detection in previous studies (Ackerman et al., 1998; S. S. Park et al., 2014). The SHAP summary plot demonstrated that the higher BTD1 contributed to estimating a higher AOD; on the other hand, the higher BTD2 led to estimating a lower AOD. In general, the AOD value increased as the BT difference between the mid-IR and thermal IR channels increased under clear-sky conditions; the SHAP and feature values of the BTD1 variable exhibited this tendency well (Lensky & Rosenfeld, 2003). In addition, aerosols such as dust usually cause negative BT difference values between 10 μm and 12 μm channels, as shown by the SHAP and feature values of BTD2 in the summary plot (She et al., 2018).

In the daytime TOA model, the differences between TOA reflectance and its 30-day minimum for each channel were identified as highly contributing variables. In particular, ch02_diff exhibited the most substantial contribution compared to the other input variables. Visible channels are known to be more sensitive to AOD than infrared channels, as they are regarded as including the effect of attenuation by scattering and absorption of aerosols and particulates in the atmosphere (He, Zha, Zhang, Gao, & Wang, 2014). Therefore, ch02_diff with a central wavelength of 511 nm contributed the most to AOD estimation, as the reference AOD data were measured at a wavelength of 550 nm. SZA was found to be more influential in the daytime TOA model than in the all-day BT model, as SZA directly affects the atmospheric radiation path (Levy et al., 2013; M. Zhang et al., 2020).

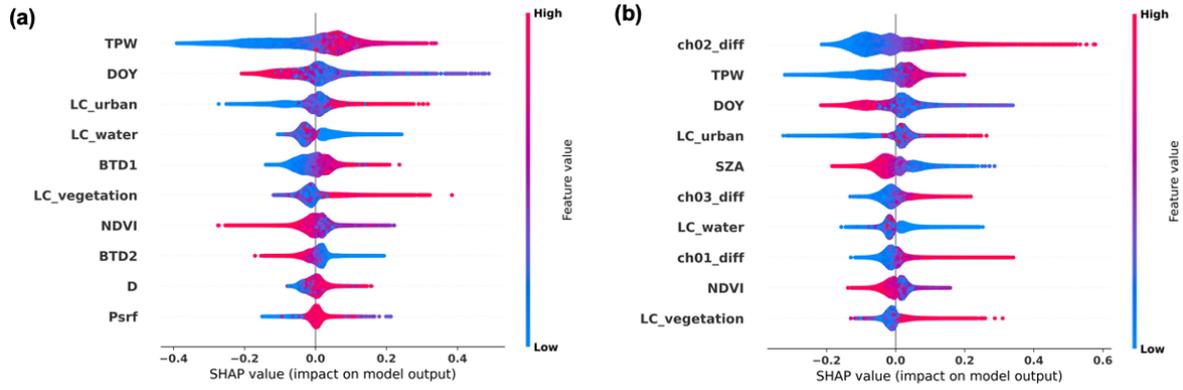

**Fig. 4.** SHAP summary plots representing the contribution of each variable in order of feature importance for (a) the all-day BT model and (b) the daytime TOA model based on LGBM. Each point on the plot indicates an individual Shapley value for a feature and its contribution to the model output. The color of each point represents the feature value, from low (blue) to high (red). Only the top 10 variables by feature importance were displayed.

### 4.3 Spatial distribution of estimated AOD

The two proposed models detected high AOD patterns in Northern China (Fig. 5), which are generally consistent with CAMS AOD. The all-day BT model estimated a slightly high AOD pattern in Southeast Asia, where CAMS and AHI YAER provide significantly high AOD patterns. The all-day BT model estimated a continuous and consistent AOD pattern for both daytime (UTC 03 and 06) and nighttime (UTC 12), even at the transition from day to night (UTC 09).

In addition, in Fig. 6, a massive aerosol transport from China to the east is detected by the proposed models and also by AHI YAER and CAMS. The high AOD distribution in China was well detected by the proposed models; however, AHI YAER and CAMS tend to estimate AOD higher than our models. During the evening and night (UTC 10, 13), the significant aerosol pattern faded compared to the past hours; however, some parts of relatively high AOD patterns still remained. Nevertheless, the proposed models demonstrated consistent AOD distribution with other AOD products in general. Especially, the all-day BT model can estimate temporally continuous distributions of AOD without interruption caused by the absence of solar reflectance at night, whereas other geostationary satellite-based AOD products are only confined to the daytime when solar reflectance exists.

Fig. 7 shows the seasonal average of the spatial distribution of AOD from the LGBM-based all-day BT and daytime TOA models with other AOD products during the daytime for test dates. A high average AOD pattern was estimated by the proposed models for central regions in China throughout all seasons. The spatial distribution of the proposed model-derived AOD exhibited similar patterns to those present in CAMS, AHI YAER, and AERONET AOD products. In particular, the all-day BT model well-detected areas with high AOD by season compared to AERONET AOD. Our study's findings on AOD distribution trends are consistent with Wang et al. (2022), showing high AOD in Southeast Asia and Southern China in spring, Northern China in summer, and low AOD annually in the Tibetan Plateau and Mongolia (P. Wang et al., 2022).

However, the seasonal average map revealed that the proposed models produced lower AOD in northern China, where both AHI YAER and CAMS AODs had high AOD distributions, despite the oversampling with AERONET AOD data in China. This is due to the lack of AERONET stations in the region, which has a high AOD tendency throughout the year, limiting sufficient training of the model. In addition, as shown in Figs. 5 and 6, the proposed models estimated the hourly spatial distribution of AOD; however, there are areas without estimation due to cloud and fog masking. It was found that inland urban areas and deserts with high AOD were largely masked out due to clouds. This is one of the limiting factors in the ability of models to estimate high AOD in large-scale urban areas in China, such as Beijing (Kang et al., 2022). Moreover, the overestimating trends in the sea of the northeastern area in spring and summer are related to the characteristics of the sea, where sea ice and sea fog frequently occur during the seasons. In particular, the Sea of Okhotsk is an area where sea fog frequently occurs in the summer, and despite fog masking using the fog detection data from GK-2A, it shows a high summer AOD average pattern due to the influence of sea fog (Sasakawa & Uematsu, 2005). Unfortunately, there are few ground observation stations in the ocean, which leads to more uncertainty in the data than on land or at sea near land with AERONET stations.

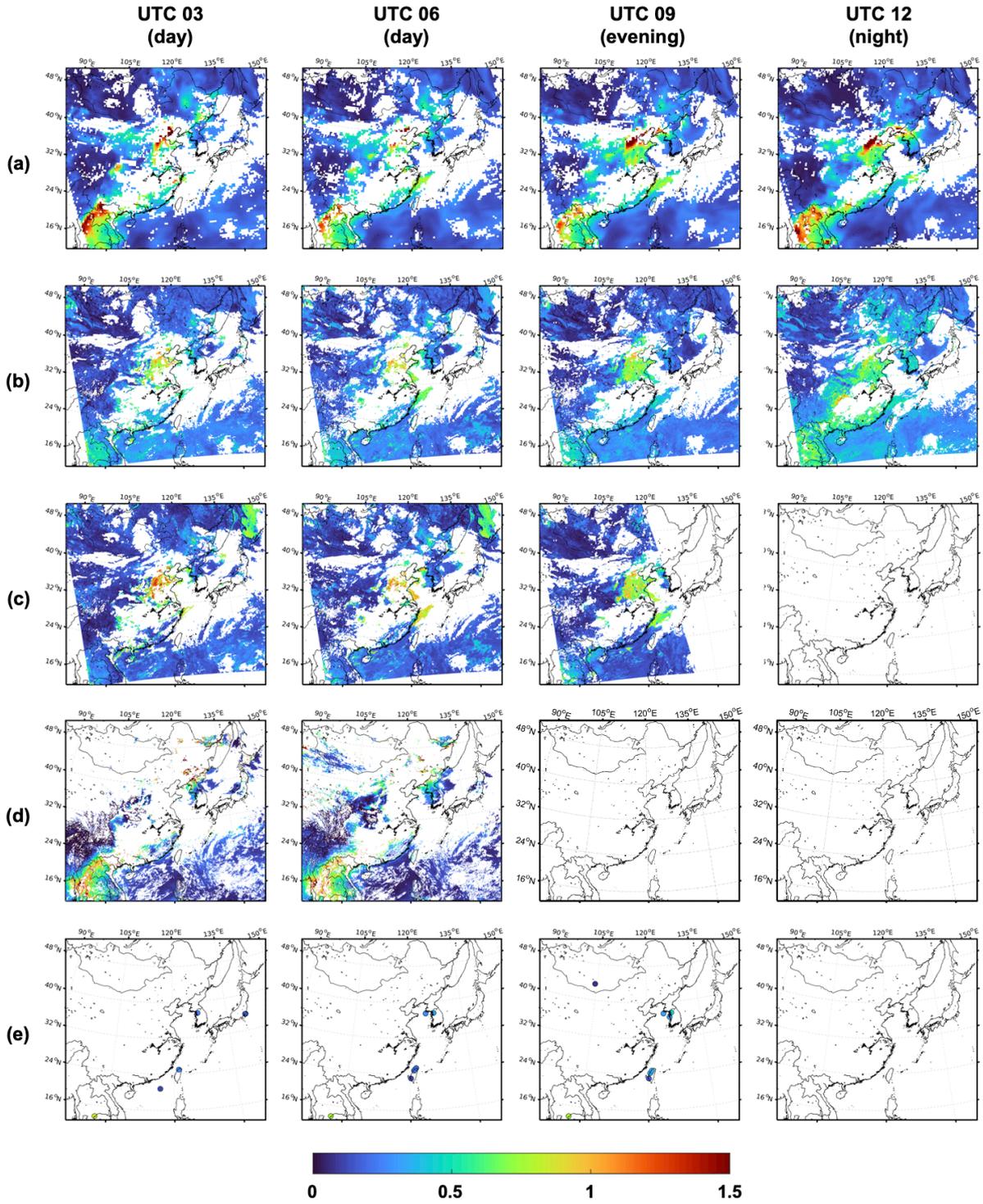

**Fig. 5.** Estimated hourly spatial distributions of AOD from the proposed models based on LGBM and other products on February 29 (DOY 60) 2020 from 3 to 12 UTC: (a) CAMS, (b) all-day BT model, (c) daytime TOA model, (d) AHI YAER, and (e) AERONET. The time and regions where AOD was not retrieved are shown as blanks. Although the CAMS forecasts gap-free AODs for the entire study area, the areas with clouds or fogs were masked out with the GK-2A L2 products for comparison with other AODs.

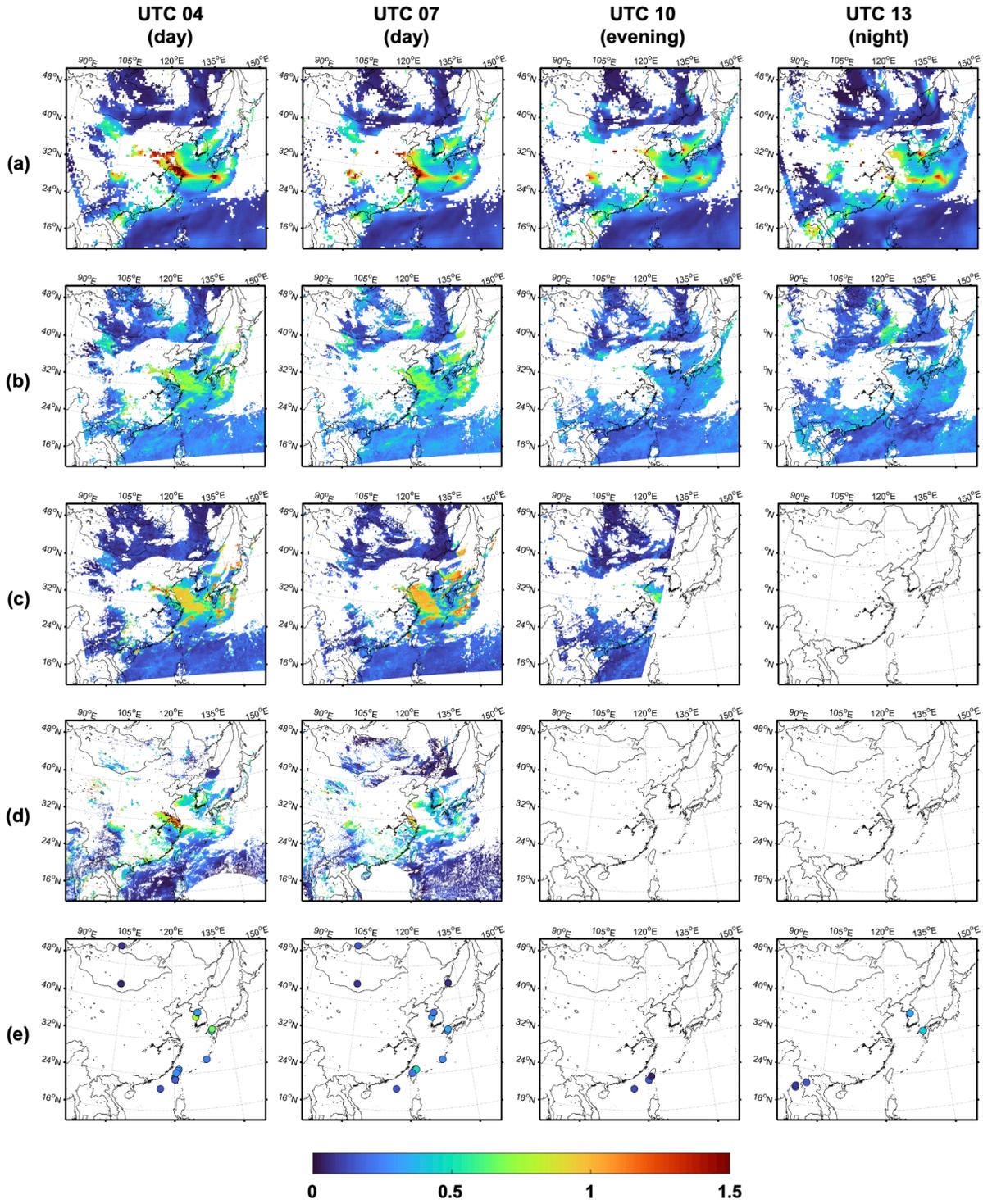

**Fig. 6.** Estimated hourly spatial distributions of AOD from the proposed models based on LGBM and other products on May 5 (DOY 125) 2020 from 4 to 13 UTC: (a) CAMS, (b) all-day BT model, (c) daytime TOA model, (d) AHI YAER, and (e) AERONET.

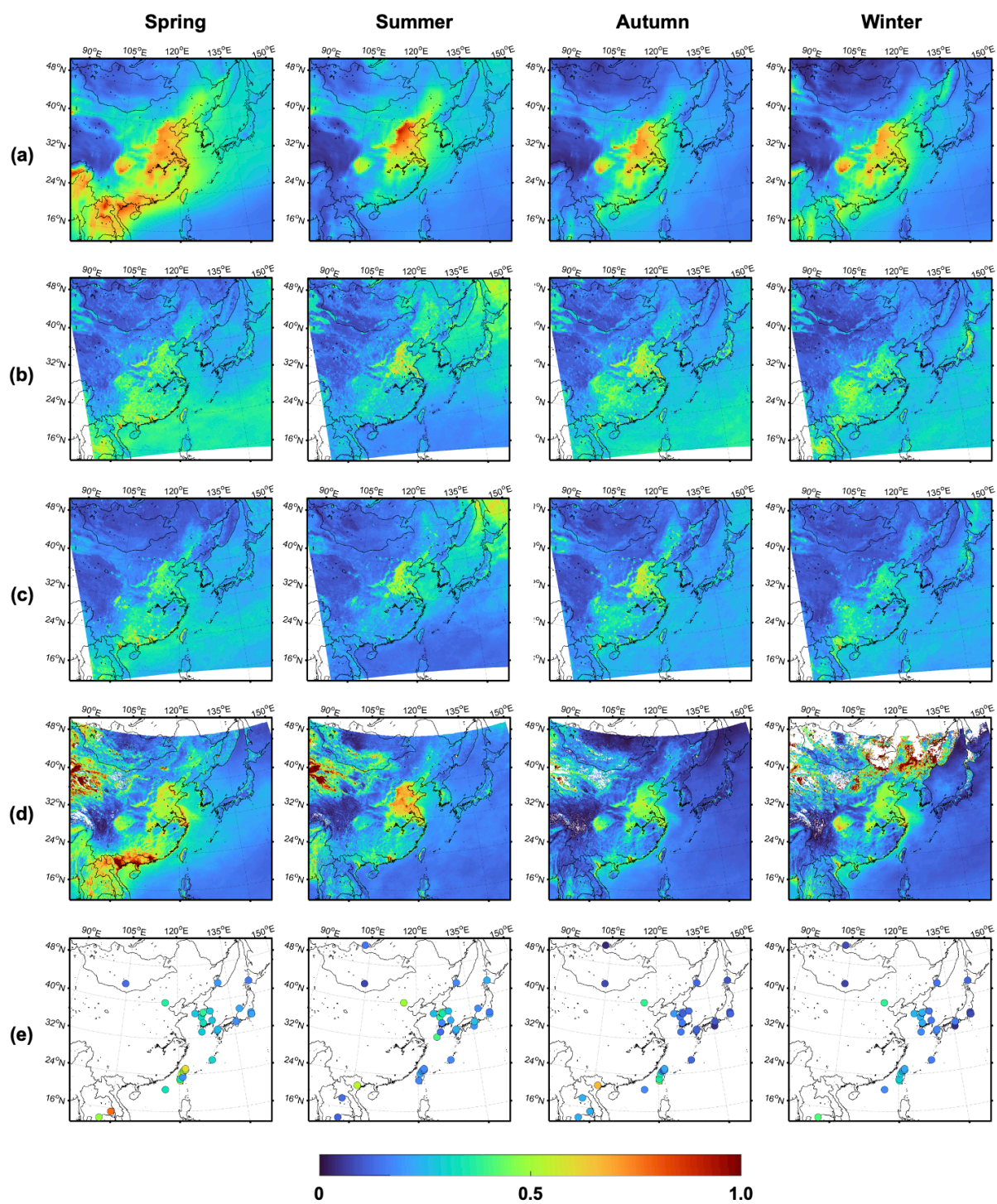

**Fig. 7.** Seasonally averaged spatial distribution of AOD estimated from (a) CAMS, (b) all-day BT model, (c) daytime TOA model, (d) AHI YAER, and (e) AERONET during the study period. All AOD data from UTC 0–7 were averaged for each season.

### 4.4 Time series analysis of estimated AOD

Overall, both proposed models exhibited similar trends with the reference AERONET AOD for high and low extremes, as well as comparable temporal patterns of fluctuation with the CAMS AOD (Fig. 8). The correlation coefficients between AODs from AERONET and the proposed models were 0.73 for the all-day BT model and 0.84 for the daytime TOA model. The correlation coefficients between AERONET and CAMS AODs were 0.77 and 0.78 for the all-day and daytime averages, corresponding with the all-day BT and the daytime TOA models, respectively. Both proposed models simulated lower AOD well; however, they also had an underestimation trend for higher AOD, as shown in the scatter plots (Figs. S3c and S3d). During June 2021, the AOD estimated by the all-day BT model and CAMS AOD had a discrepancy with the AERONET AOD; on the other hand, the daytime TOA model had a pattern similar to the AERONET AOD. In July 2021, there were not enough clear days to estimate AOD due to the East Asian monsoon, which caused discrepancies among AODs from AERONET, CAMS, and proposed models. Nonetheless, the proposed model overall simulates the daily averaged time series pattern similar to the AERONET AOD.

During a period of one week, both proposed models simulated the hourly variation trend of AOD, which closely resembled that of the reference AERONET AOD (Fig. 9). In addition, they had temporal patterns comparable with CAMS AOD. The all-day BT model exhibited more continuous diurnal variation of AOD every hour with noticeable ascending and descending trends as it provided both daytime and nighttime estimations. Therefore, it was confirmed that the all-day BT model estimated AOD continuously throughout the day and night, similar to the reference AERONET AOD.

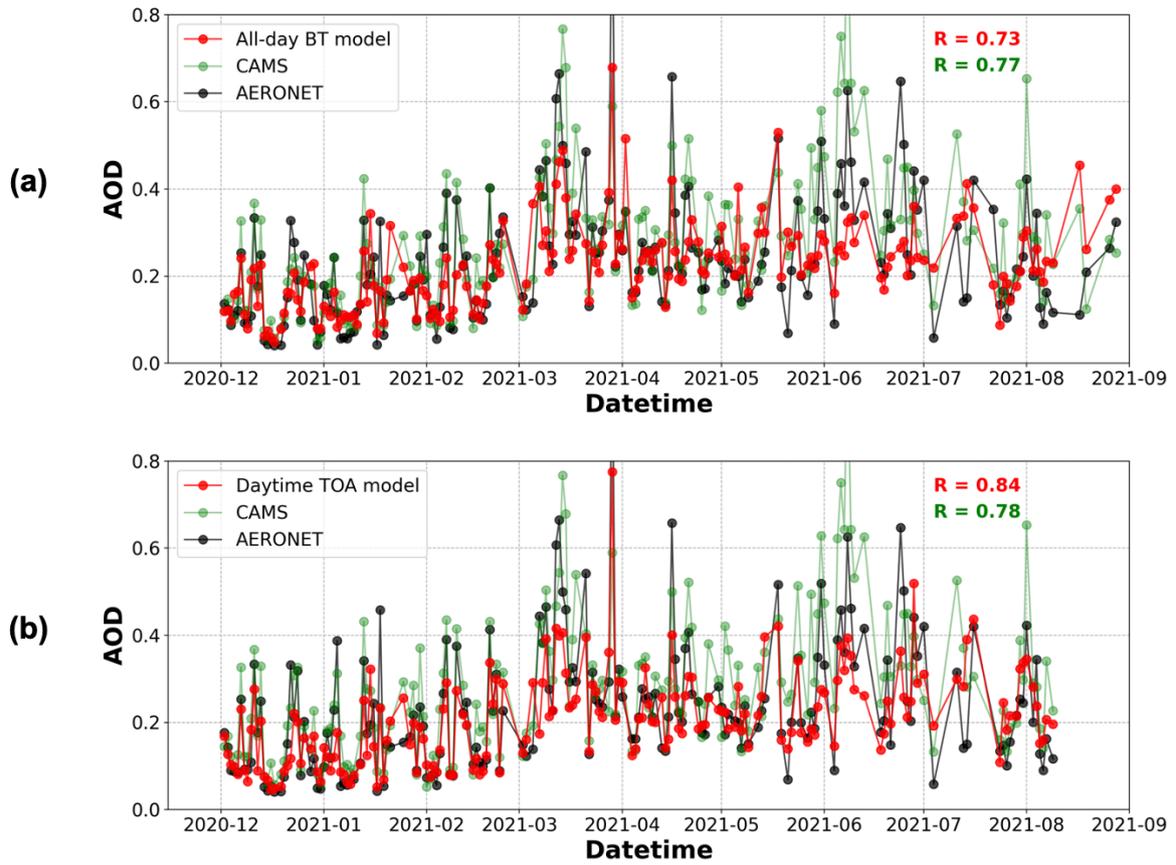

**Fig. 8.** Time series of daily averaged estimated AOD (red) from (a) the all-day BT model and (b) the daytime TOA model based on LGBM in comparison to the AERONET (black) and CAMS AODs (green) for test station (KORUS_UNIST_ULSAN) from December 2020 to August 2021.

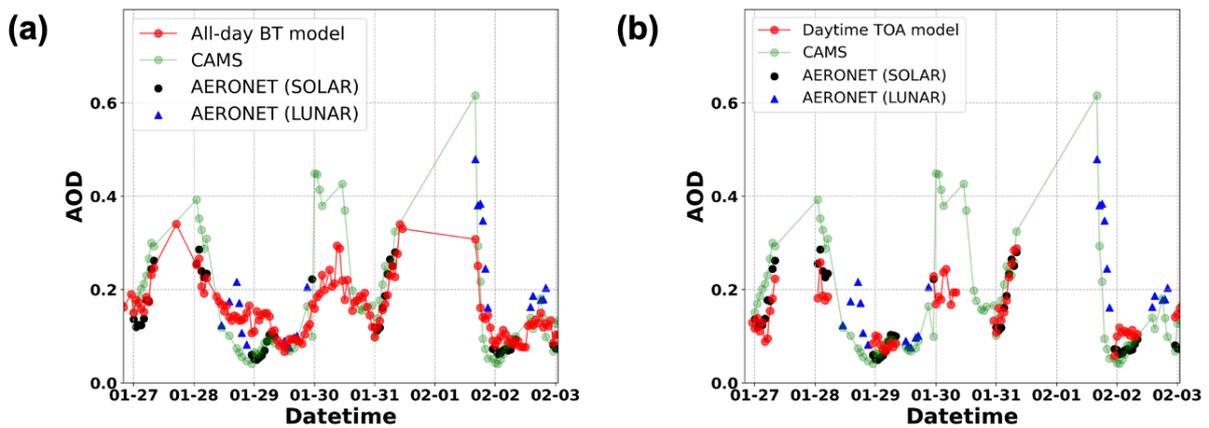

**Fig. 9.** Time series of hourly estimated AOD (red) from (a) the all-day BT model and (b) the daytime TOA model based on LGBM in comparison to the AERONET (black for solar AOD and blue for lunar AOD) and CAMS AODs (green) for test station (KORUS_UNIST_ULSAN) from January 27 to February 2, 2021.

### 4.5 Novelty and limitations

Although operational geostationary satellite-derived AOD products cover large areas, they are not available at night. The proposed all-day BT model using machine learning enables the estimation of AOD during the night, which is a novel aspect of this research. Consequently, it is possible to continuously monitor the aerosol transport and diurnal variability of AOD every hour, which was impossible with existing algorithms using geostationary satellites. Furthermore, the spatial resolution of the estimated AOD from the proposed models is better than those of the currently available nighttime AOD products from reanalysis or forecast data. For example, CAMS global atmospheric composition forecasts (0.4° × 0.4°, approximately 40 km), CAMS global reanalysis (0.75° × 0.75°, approximately 75 km), and the second Modern-Era Retrospective analysis for Research and Applications (MERRA-2; 0.5° × 0.625°, approximately 50 km in the direction of latitude) provide AOD with coarser spatial resolutions than the proposed models (ECMWF, 2021; Gueymard & Yang, 2020).

However, some limitations exist in this study: First, the number of AERONET stations was limited, and the available reference data were unevenly distributed on both the spatial and temporal domains. Especially, the AERONET stations are sparse in China, Mongolia, and the open ocean. The limited number of AERONET stations often leads to high uncertainty in the areas. In addition, the number of AERONET AOD data between day and night is different. Different from the solar AOD during the daytime, the lunar AOD at night is affected not only by clouds but also by the phase of the moon because the time of the rise and set of the moon and the intensity of lunar radiation change according to the moon phase. Second, the AOD estimation is unavailable under cloudy conditions, which is a common limitation for satellite-based aerosol products. Finally, there are no operating satellite-based nighttime AOD products for comparison. Since this study is the first to estimate nighttime AOD over a wide area using geostationary satellite observations, the only available data to compare are AOD forecast or reanalysis products.

### 5. Conclusion

In this study, we estimated temporally continuous AOD at 550 nm during both daytime and nighttime using GK-2A BT data by machine learning-based methods. The estimated AOD from the proposed models generally agrees well with the ground-based AERONET AOD. For both all-day BT and daytime models, the most commonly contributing input variables were both TPW and DOY. For the all-day BT model, the BT differences turned out to be the following important variables, and for the daytime TOA model, on the other hand, the additionally used TOA reflectance differences were found to be significant variables. The spatial distributions of estimated AOD are generally in good agreement with the corresponding AHI YAER, CAMS, and AERONET AOD in one of the test days. Both proposed models show underestimating trends in spatial distribution; nevertheless, the spatial distributions of regions with high AOD of seasonal averaged AOD are generally in good agreement with other AOD products. The time series of AOD at the test station estimated from the proposed models are consistent with AERONET and CAMS AOD on an hourly scale and daily average.

This study demonstrated that geostationary satellite-based AOD estimation, including nighttime, is possible with machine learning approaches. It is expected that continuous monitoring of the transport of aerosols in East Asia and the diurnal variability of AOD in a specific area is possible using the temporally continuous AOD estimation through the model proposed in this study. The estimated AOD from the proposed models, including nighttime AOD, can be used as a parameter for satellite-based PM retrieval algorithms as well as a key factor for estimating the radiative forcing of aerosols for the study of climate change.

Supplementary

**Table S1.** Channel information of GK-2A AMI

| Channel number | Channel type | Central wavelength (μm) | Spatial resolution (km) |
|---|---|---|---|
| 1 | Visible | 0.470 | 1 |
| 2 | Visible | 0.511 | 1 |
| 3 | Visible | 0.640 | 0.5 |
| 4 | Visible | 0.856 | 1 |
| 5 | Near-IR | 1.380 | 2 |
| 6 | Near-IR | 1.610 | 2 |
| 7 | Shortwave-IR | 3.830 | 2 |
| 8 | Mid-IR | 6.241 | 2 |
| 9 | Mid-IR | 6.952 | 2 |
| 10 | Mid-IR | 7.344 | 2 |
| 11 | Thermal-IR | 8.592 | 2 |
| 12 | Thermal-IR | 9.625 | 2 |
| 13 | Thermal-IR | 10.403 | 2 |
| 14 | Thermal-IR | 11.212 | 2 |
| 15 | Thermal-IR | 12.364 | 2 |
| 16 | Thermal-IR | 13.310 | 2 |

**Table S2.** Channel information of GK-2A AMI

| Algorithm | Hyperparameter | Definition | Value |
|---|---|---|---|
| RF | n_estimators | number of trees | 500 |
|  | max_depth | maximum depth of the tree grown | 20 |
|  | max_features | maximum number of features to consider for the best split at each node | the square root of the number of features |
|  | criterion | quality of a split | the mean squared error |
| LGBM | boosting_type | boosting type | not fixed |
|  | num_leaves | maximum number of leaves |  |
|  | n_estimator | number of boosted trees to fit |  |
|  | learning_rate | boosting learning rate |  |
|  | max_depth | maximum tree depth |  |
|  | min_child_weight | minimum sum of instance weight needed in a child leaf |  |
|  | subsample | subsample ratio of the training instance |  |
|  | colsample_bytree | subsample ratio of columns when constructing each tree |  |

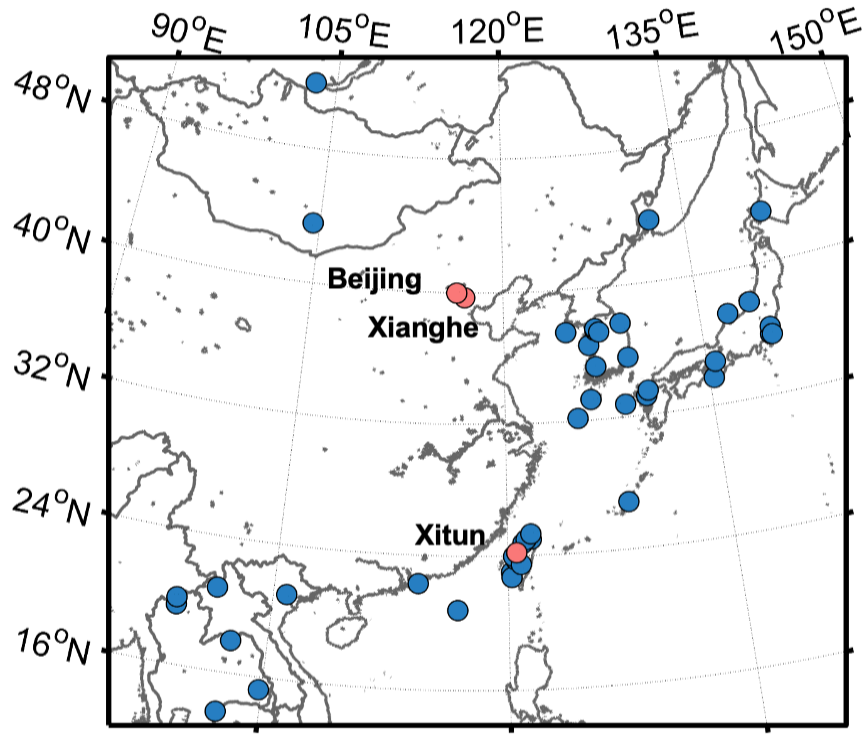

**Fig. S1.** The distribution of AERONET stations for training and testing in the study area. The stations with oversampling applied are marked as red circles and the remaining stations without oversampling are marked as blue circles.

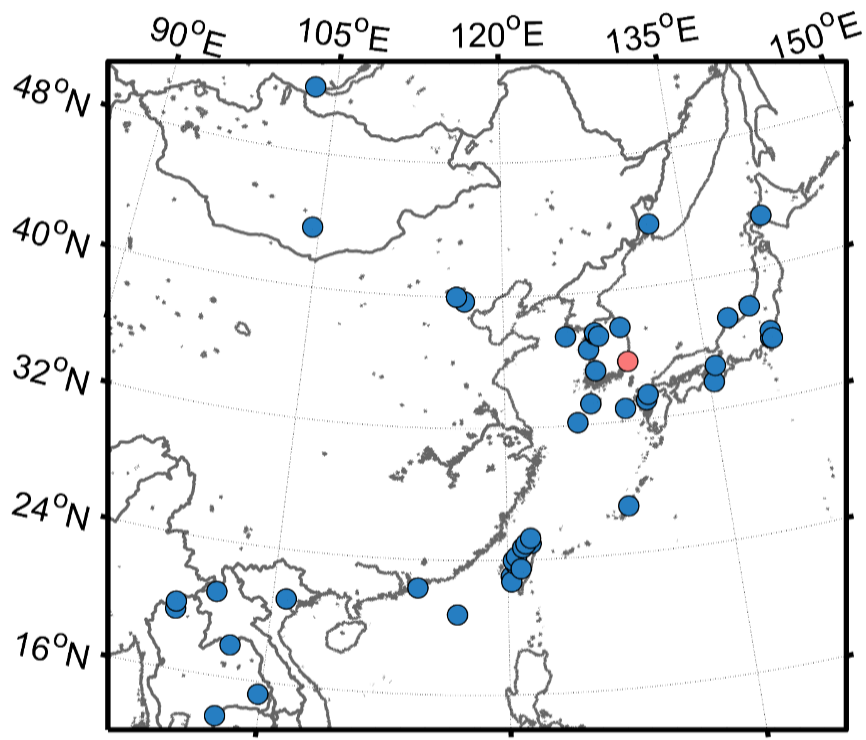

**Fig. S2.** The distribution of AERONET stations for training and testing in the study area. The training stations are marked as blue circles and the test station is marked as a red circle.

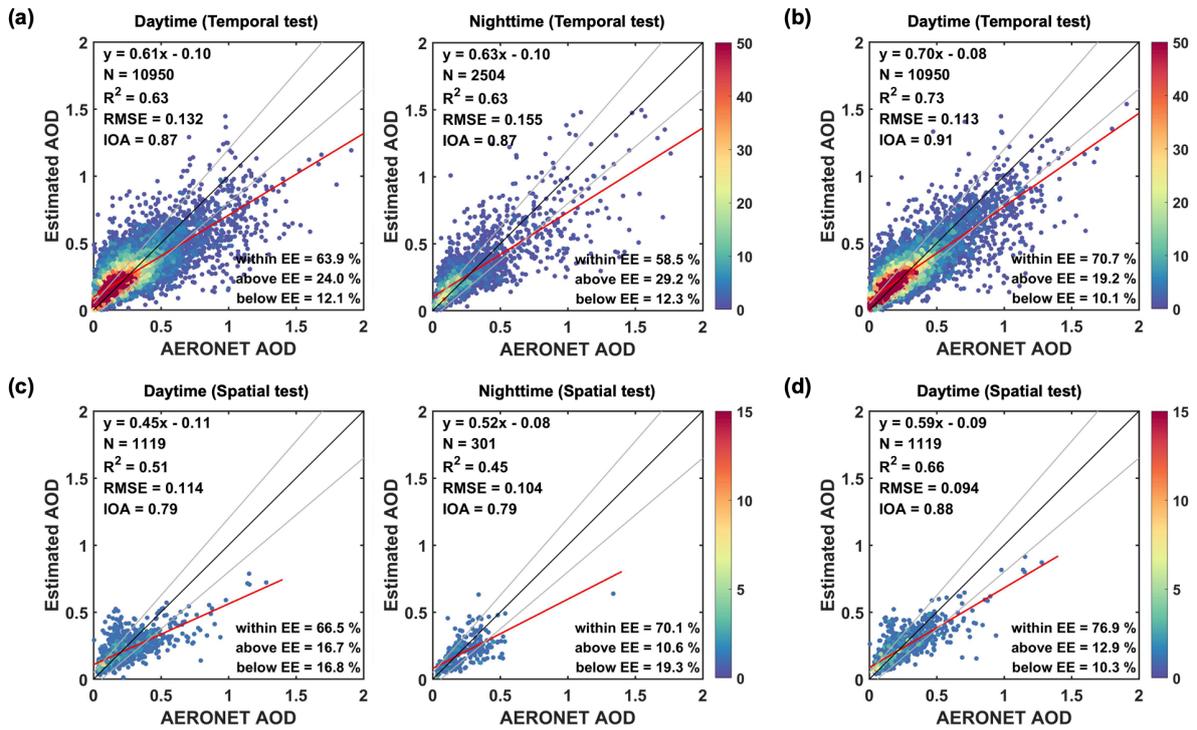

**Fig. S3.** Scatter plots of the temporal for (a) the all-day BT and (b) the daytime TOA models, and the spatial test for (c) the all-day BT model and (d) the daytime TOA model based on LGBM.

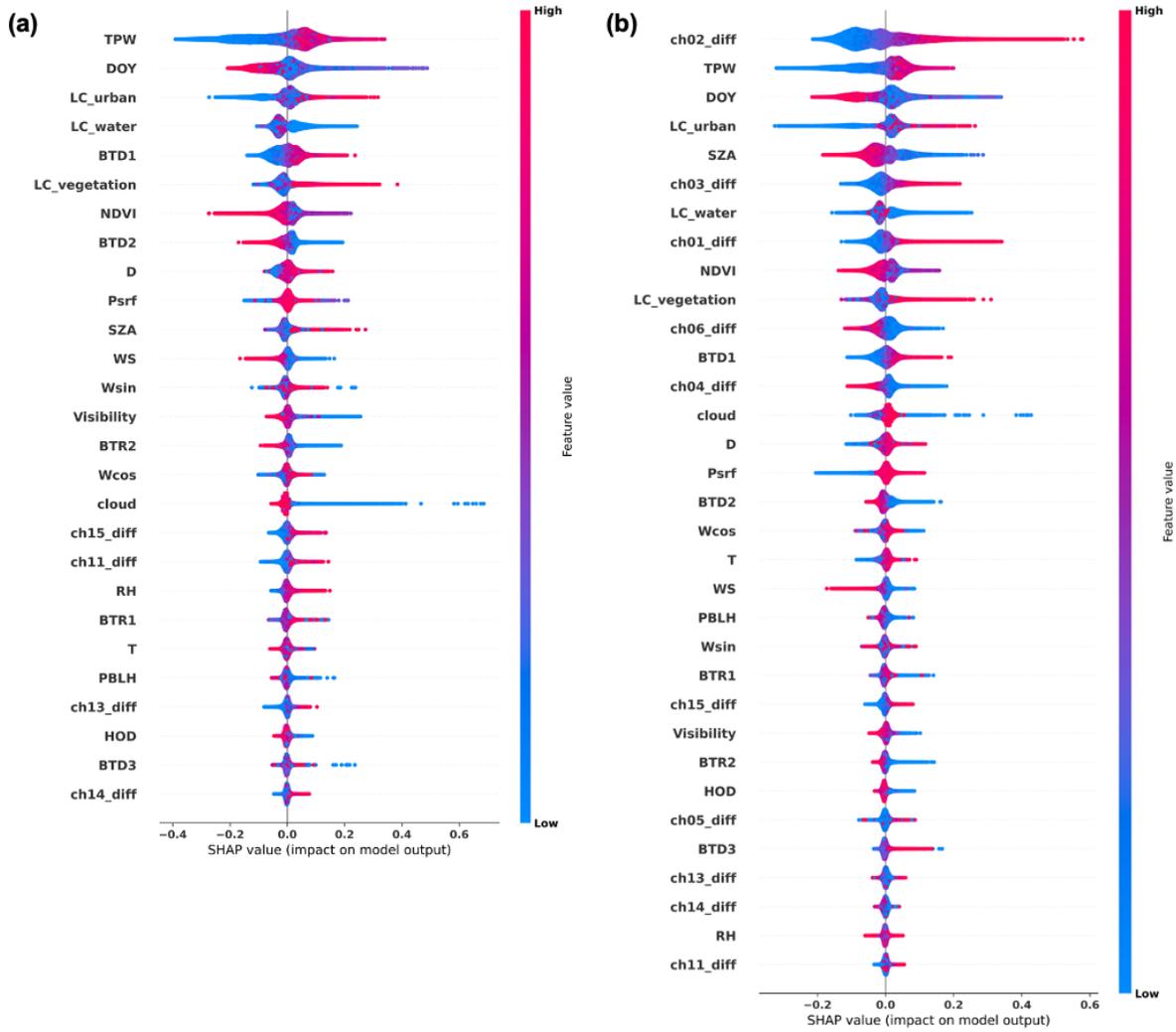

**Fig. S4.** SHAP summary plots with feature importance and contribution for entire input variables for (a) the all-day BT and (b) the daytime models for LGBM.